\documentclass[letterpaper,twocolumn,10pt]{article}
\usepackage{usenix,epsfig,endnotes}

\usepackage{cite}
\usepackage{amsmath,amssymb,amsfonts}
\usepackage{algorithmic}

\usepackage[utf8]{inputenc}
\usepackage[english]{babel}
\usepackage{indentfirst}

\usepackage{dblfloatfix}
\usepackage{adjustbox}
\usepackage{lipsum}
\usepackage{placeins}
\usepackage{graphicx}
\usepackage{textcomp}
\usepackage{colortbl}
\usepackage{multirow}
\usepackage{booktabs}
\usepackage{amssymb}
\usepackage{pifont}
\usepackage{tcolorbox}
\def\BibTeX{{\rm B\kern-.05em{\sc i\kern-.025em b}\kern-.08em
  T\kern-.1667em\lower.7ex\hbox{E}\kern-.125emX}} 

\begin{document}

\date{}

\title{\Large \bf TANTRA: Timing-Based Adversarial Network Traffic Reshaping Attack}
\author{
{\rm Yam Sharon}\\
Ben Gurion University of the Negev
\and
{\rm David Berend}\\
Nanyang Technological University
\and
{\rm Yang Liu}\\
Nanyang Technological University
\and
{\rm Asaf Shabtai}\\
Ben Gurion University of the Negev
\and
{\rm Yuval Elovici}\\
Ben Gurion University of the Negev
}

\maketitle

\begin{abstract}
Network intrusion attacks are a known threat.
To detect such attacks, network intrusion detection systems (NIDSs) have been developed and deployed. 
These systems apply machine learning models to high-dimensional vectors of features extracted from network traffic to detect intrusions.
Advances in NIDSs have made it challenging for attackers, who must execute attacks without being detected by these systems. 
Prior research on bypassing NIDSs has mainly focused on perturbing the features extracted from the attack traffic to fool the detection system, however, this may jeopardize the attack's functionality.
In this work, we present TANTRA, a novel end-to-end  \textbf{T}iming-based \textbf{A}dversarial \textbf{N}etwork \textbf{T}raffic \textbf{R}eshaping \textbf{A}ttack that can bypass a variety of NIDSs.
Our evasion attack utilizes a long short-term memory (LSTM) deep neural network (DNN) which is trained to learn the time differences between the target network's benign packets. 
The trained LSTM is used to set the time differences between the malicious traffic packets (attack), without changing their content, such that they will "behave" like benign network traffic and will not be detected as an intrusion. 
We evaluate TANTRA on eight common intrusion attacks and three state-of-the-art NIDS systems, achieving an average success rate of 99.99\% in network intrusion detection system evasion.
We also propose a novel mitigation technique to address this new evasion attack.
\end{abstract}

\section{Introduction}
% network intrusion
The networks of governments, enterprises, and private households are the target of network intrusion attacks performed by attackers with an interest in financial gain, disruption, or gathering intelligence. 
According to recent trends, the most prominent reason for targeted network intrusion attacks is acquiring intelligence. Such attacks increased tenfold from 2017 to 2018~\cite{symantec_2019}. 
In order to launch an attack on an organizational network (for example, a man-in-the-middle (MITM) attack), the attacker must first bypass the targeted network's defense mechanisms.

% defense
The defense mechanisms aimed at thwarting such attacks are referred to as network intrusion detection systems (NIDSs); to address the increasing complexity and sophistication of attacks, these systems have become more advanced in recent years.
At the core of these defense advancements lies the extraction of high-dimensional vectors of features from network traffic, which are then utilized by traditional machine learning or deep learning models trained to distinguish between benign and malicious traffic. 

% feature space related work
Advances in NIDSs have presented attackers with a challenge which they have shown limited ability in overcoming~\cite{chalapathy2019deep, wang2019survey, 4580100}. The majority of evasion attack approaches are based on perturbing the attack's traffic features by assuming a white-box setting in which the attacker has knowledge about the system's feature extractor. This knowledge is then used to perturb malicious traffic by performing an adversarial attack aimed at bypassing the NIDS. 
While shown to be effective at bypassing the NIDS, these attacks lose some of their original functionality after the adversarial perturbation is applied. Thus, the attacks' impact on the target network remains limited, as the attacks may not be viable in an end-to-end setting.

% timing related work
Given the limitations of feature-based attacks, new ways of evading NIDSs are likely to be explored by attackers. One promising direction aims at enabling malicious intrusion traffic to mimic benign network behavior.
In one of the first works aimed at mimicking benign traffic, Han et al.  reshaped the timing of malicious network traffic and added dummy packets to avoid detection by the NIDS~\cite{han2020practical}. This approach requires preparing all of the attack traffic on the attacker's side and was not evaluated in an end-to-end scenario. Instead, the performance was measured against a NIDS, without setting up an active network connection. Given the need to prepare all of the attack traffic on the attacker's side in advance, the approach cannot be adjusted according to frequent target network responses or delays. This makes it less applicable for real-world settings.

% our approach
In this paper, we present a new \textbf{T}iming-based \textbf{A}dversarial \textbf{N}etwork \textbf{T}raffic \textbf{R}eshaping \textbf{A}ttack (TANTRA) that mimics benign traffic to evade detection. TANTRA's main advantages are that it maintains the attack packets' content and can be adjusted according to frequent target network responses in real time.
Furthermore, no knowledge on the NIDS is required, resulting in a fully black-box attack. This makes TANTRA the first successful real-world evasion attack against NIDSs. 

The proposed evasion attack is based on training a long short-term memory (LSTM) to generate the time differences between benign network packets (collected from the target network in advance). Then, the trained model is used to reshape the malicious traffic's inter-packet delay. This way, the modified malicious traffic is able to mimic the behavior of benign traffic.
We evaluate TANTRA on eight common network attacks against three state-of-the-art (SOTA) NIDSs that are based on traditional machine learning (ML) and deep learning (DL) techniques. Our novel evasion attack successfully evaded all three defenses, achieving a 99.99\% success rate (on average), outperforming evasion attacks proposed in prior research under similar settings by up to 28.8\% (as shown in Section \ref{sec:evaluation}).

% contributions
In summary, our contributions are as follows:
\begin{itemize}
    \item We present TANTRA, a novel network evasion attack that can evade SOTA defenses with a 99.99\% success rate. The proposed attack outperforms methods/attacks proposed in related work and performed under similar conditions by up to 28.8\% more effectively.
    \item TANTRA can evade NIDSs without changing the attack packets' content, eliminating the risk of losing the attack's functionality, which has been a prominent drawback in related work.
    \item TANTRA is the first end-to-end evasion attack against NIDSs that operates in a black-box setting, as successfully demonstrated against three SOTA NIDSs in this paper. 
    \item To mitigate the threat posed by this new evasion attack, we propose a novel defense technique. A preliminary version of this mitigation was implemented, evaluated, and shown to be promising.
\end{itemize}

% structure
The remainder of the paper is structured as follows. In Section \ref{sec:background}, the background is presented; this is followed by the threat model, which is described in Section \ref{sec:threatmodel}. The attack methodology is discussed in Section \ref{sec:methodology}, followed by a discussion of the weaknesses and mitigation of NIDSs in Section \ref{sec:theorydefense}. We present the evaluation in Section \ref{sec:evaluation} and provide a detailed comparison to related work in Section \ref{sec:comparison}. Our conclusions are presented in Section \ref{sec:conclusion}. 

\begin {table*}[t!]
\center
\caption {Comparison to prior research.}
\label{tab:approachrequirements} 

\resizebox{0.55 \textwidth}{!}{%
\begin{tabular}{ccc}
\hline\hline
Assumptions & Related work & Ours \tabularnewline
\hline\hline
Ongoing connection to target network & \checkmark \cite{Rallying, adv04, IDSGAN, IDSVol, DLNIDS, adv03} & \checkmark\tabularnewline
Knowledge on feature extractor & \checkmark \cite{Rallying, adv04, IDSGAN, IDSVol, DLNIDS, adv03} & .\tabularnewline
Ability to change extracted features  & \checkmark \cite{Rallying, adv04, adv03}& .\tabularnewline
Access to NIDS & \checkmark \cite{Rallying, adv04, adv03}& .\tabularnewline
\hline\hline
\end{tabular}
}
\vspace{0.7em}
\end {table*}

\section{Background}\label{sec:background}

\subsection{Network Intrusion Detection Systems}

% network intrusion attacks
Network intrusion attacks refer to any unauthorized activity on a target network.
To detect and prevent such attacks, NIDSs are  employed. These systems are based on two main approaches: signature-based and anomaly-based approaches~\cite{buczak2015survey}. 
The signature-based approach is a lightweight defense where the defender manually defines a set of rules that characterize malicious traffic. This approach is easy to implement but is also easy to bypass, as the attacker can determine the rules and reshape the malicious traffic in light of the rules defined by the defender. It becomes more difficult for the attacker when the anomaly-based approach is utilized. When combined with deep learning techniques, this defense has been shown to be very effective~\cite{chalapathy2019deep, kwon2019survey}. Together with the high-dimensional features extracted from the traffic, the NIDS receives a fine-grained perspective on the traffic characteristics; therefore, the NIDS can more effectively detect incoming attacks as anomalous. Hence, reshaping the malicious traffic so that it will not be detected as anomalous is challenging.

\subsection{Traffic Reshaping} 

% basic concept
Given the effectiveness of recent NIDSs, attackers will be forced to explore new ways of performing network intrusion attacks. To be successful, the network intrusion attacks require adjustments to prevent detection by any NIDS. Modifying the network traffic packets of the attack is referred to as traffic reshaping; there are several approaches used for reshaping. One approach is to reshape the features extracted from the network traffic (feature-space reshaping). Another emerging direction is to reshape the timing of malicious network packets (timing-based reshaping).

For feature-space reshaping, attackers require knowledge of the target network feature extractor and NIDS, which is difficult to obtain.
In this case, perturbations are applied to the extracted features in order to bypass the NIDS. These perturbations are specifically crafted for the targeted NIDS. The perturbations are referred to as adversarial examples, which, in the context of network intrusion, result in adversarial attacks.
% origin white-box & rationale
Adversarial attacks are inspired by the computer vision (CV) domain, where Szegedy et al. \cite{szegedy2013intriguing} first created a specially crafted pixel mask invisible to the human eye, causing the neural network to misclassify the input image. Later, Goodfellow et al. \cite{goodfellow2014explaining}, Carlini and Wagner \cite{carlini2017towards}, and others~\cite{papernot2016limitations, chen2017ead} advanced the field by minimizing the adjustments needed
for a white-box setting. For intrusion attacks, such a setting requires extensive knowledge of the preprocessing techniques and behavior of the targeted NIDS.
Previous studies have demonstrated successful intrusions when assuming a white-box setting~\cite{IDSGAN, adv03,pierazzi2019intriguing, adv04, IDSVol, DLNIDS, Rallying}; however when a white-box setting is required, applicability in the real world is minimal.
For greater real-world applicability, a black-box setting (commonly referred to as an end-to-end scenario) is required. For this scenario, knowledge and access to the targeted network's feature extractor and feature space are not available.
Furthermore, for the end-to-end scenario, the attacker must be capable of adjusting the evasion attack according to frequent responses of the target network.
Finally, the attack's functionality must be maintained.

Timing-based reshaping is emerging as an approach that does not require any knowledge of the NIDS. Its only requirement is the need to obtain benign traffic from the target network. This makes it a suitable candidate for a black-box setting. In~\cite{han2020practical} the authors used a generative adversarial network (GAN) to combine dummy packets with the malicious network packets to change the timing of the attack's traffic. While showing promise for a scenario in which there is only limited knowledge on the target network, an end-to-end scenario was not able of being performed. % the authors were unable to demonstrate a full end-to-end scenario.
To accomplish this, one remaining challenge that needs to be addressed lies in its methodology, where target network responses cannot be integrated. So far, the attack traffic is prepared (in its entirety) in advance which prevents response integration. Furthermore, the additional dummy packets pose a risk to the attack's functionality. We aim to address such drawbacks by demonstrating a full end-to-end attack scenario without risking the attack's functionality.

\section{Threat Model}\label{sec:threatmodel}

% Goal
In this study, we assume that (1) the attacker treats the NIDS as a black-box, (2) the attacker has access to the target network, enabling him/her to observe the benign traffic (to learn its patterns) and send the reshaped attack traffic to the network,
(3) the attacker adjusts the malicious traffic in real time, and 
(4) the attacker maintains the malicious packets’ content.
These assumptions represent a realistic threat model and to the best of our knowledge, were not addressed in prior work.

During the setup phase, the attacker learns the benign timing behavior of the target's network traffic. 
Then, the malicious network traffic is reshaped in real time, in accordance with the learned benign timing behavior, so that the target NIDS will not issue an alert.
One of the main strengths of the proposed threat model is the limited assumptions necessary for its execution, as presented in Table~\ref{tab:approachrequirements}, which compares our assumptions to those of prior studies~\cite{IDSGAN, adv03, adv04, IDSVol,DLNIDS, Rallying}. 
In our approach, the attacker does not need access to the targeted NIDS or knowledge about its behavior (e.g., extracted features, NIDS type, architecture). 
The only requirement for the attacker is an ongoing connection to the target network, a requirement for any network intrusion attack. 

\begin{figure*}[t]
\begin{center}
    \includegraphics[width=1.0\textwidth]{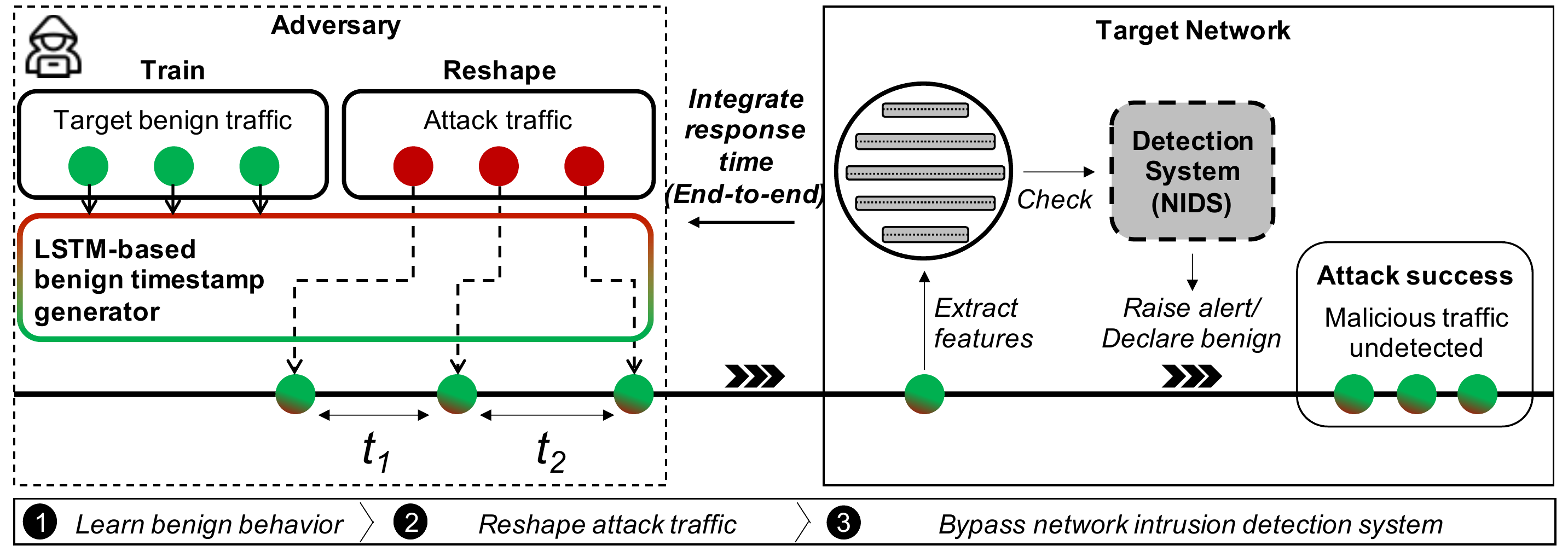}
    \caption{Overview of our evasion attack.}
    \label{fig:attack_overview}
\end{center}
\end{figure*}

\section{TANTRA: Novel Evasion Attack}\label{sec:methodology}

\subsection{Main Challenges} \label{sec:mainchallenges}

An effective evasion attack against a NIDS was demonstrated in prior research, however, there were difficulties in retaining the attack's functionality, and a white-box setting was required.
The methodology used in our study reflects our aim of crafting an attack capable of both maintaining the attack packets' content and evading the NIDS, with limited available knowledge. To achieve this we have to overcome two main challenges. The first challenge is switching the reshaping context from the commonly used feature space to the less-explored problem space. The second challenge is utilizing the limited network characteristics in the problem space for an effective evasion attack in a black-box setting. 

When analyzing other approaches proposed for evading NIDSs~\cite{adv03,pierazzi2019intriguing, adv04, IDSVol, DLNIDS, Rallying}, we see that retaining the functionality of the attack packets' content does not seem possible.
In fact, Pierazzi et al.~\cite{pierazzi2019intriguing} found that modifying the feature space after feature extraction may lead to a loss in functionality. 
Analyzing their results, we found that extracted features are based on statistical functions, such as the mean and STD, which are difficult (if not impossible) to reverse once a perturbation has been added. 

Furthermore, the extracted features are related to each other and calculated from the same network packets’ attributes. Changing one feature without changing the other related high-dimensional features will make it even more problematic to reverse the traffic to its original form. Therefore, we conclude that the feature space does not seem like a promising avenue to pursue for crafting an effective attack. Instead, we turn to the problem space. Here, reshaping can be done before the features are extracted from the traffic by the NIDS, and therefore it provides a foundation for retaining the functionality of the attack in a black-box setting.

However, the problem space offers limited options for reshaping the network traffic characteristics. For example, most of the time, the attacker is unable to control or modify the network connection characteristics, such as the destination's MAC and IP addresses. Two of the limited modifiable options are the packet size and~timestamp features; the attacker can split or pad the packets' content, and he/she can send the packets with different timestamps. In this work, we show that successful evasion can be achieved by reshaping just the packet timestamps, without influencing the packet's size.

\subsection{Overview}

\begin{figure*}[t!]
    \centering
    \includegraphics[width=0.70\textwidth]{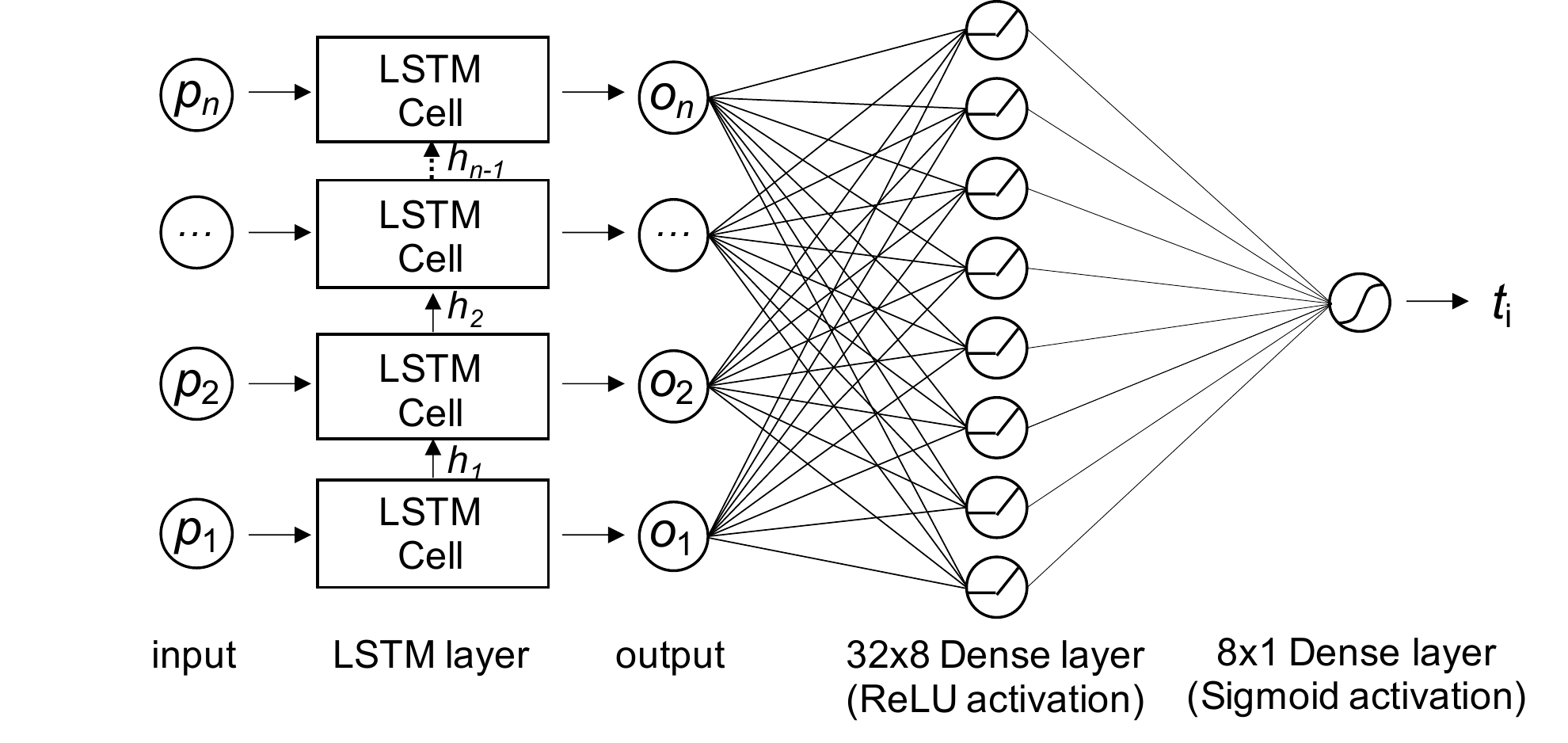}
    \caption{Overview of the DNN model architecture.}
    \label{fig:LSTM}
\end{figure*}

Our evasion attack consists of three steps, which are illustrated in Figure~\ref{fig:attack_overview}.
In step \ding{182}, an LSTM is trained to learn benign network traffic behavior by predicting the time differences between benign network packets.
The LSTM is trained for the specific network connection on which the attack will be executed.
In step \ding{183}, the trained LSTM is applied on an intrusion attack's malicious traffic to reshape the interpacket delay (adjusting the timestamps), so it is similar to that of benign traffic.
In step \ding{184}, the reshaped malicious traffic is sent to the target network where it aims to bypass the NIDS which is treated as a black-box. 

\subsection{Preparation}
To reshape malicious traffic so it behaves like benign traffic, we utilize a many-to-one LSTM model. The LSTM's input is a short history of previous packets' timestamps, along with the size of the previous and current packets. To determine the optimal amount of history of network packets (window size), we perform an empirical assessment (see Section \ref{sec:evaluation}), starting with a window size (WS) of three and increasing it to 150.
The LSTM's output is the timestamp for the current network packet. The architecture (Figure \ref{fig:LSTM}) consists of an LSTM cell for each input packet $p$, with a hidden layer $h$ and cell output $o$. The cell output is passed on to a $32x8$ dense layer with ReLU activation. Finally, an 8x1 dense layer with sigmoid activation generates the timestamp $t$ (this architecture is based on previous work that proposed a many-to-one LSTM \cite{roondiwala2017predicting, cao2019financial}).
\begin{equation}
f(history, p_{i}) = t_{i}
\end{equation}

During the training phase, the LSTM receives the benign traffic collected from the targeted network. In each training iteration, the LSTM receives $n$ packets as input. Afterward, $n$ timestamp predictions are compared with the known interpacket delays to compute the mean squared error (MSE) loss. The loss is then backpropagated through the LSTM. 
After the training phase, the LSTM is assumed to be capable of predicting interpacket delays between the current and previous packets using timestamps. 
One advantage of the LSTM architecture is its sequential behavior, where the traffic is processed packet by packet. This enables an end-to-end scenario.

\subsection{Execution} \label{sec:meth:apply}

The trained LSTM can be utilized to predict the timestamps of the attack, packet by packet. This enables an end-to-end scenario in which target network delays and responses are integrated during the attack reshaping process. 

Since TANTRA is the first evasion attack capable of performing an end-to-end scenario, comparing the performance in this setup to related work may be inaccurate. Hence, we first analyze its performance directly against the NIDS, without an active target network connection, following the setup of prior research~\cite{Rallying, adv04, IDSGAN, IDSVol, DLNIDS, adv03}.
In this scenario, malicious network traffic stemming from, e.g., an MITM attack, is reshaped by adjusting the timestamps and evaluated directly against the NIDSs.

For the end-to-end scenario, the timestamps are generated step by step with an active target network connection. Here, the MITM attack is not directly evaluated on the NIDS; instead, it is sent in multiple steps to the target network. When the packets reach the target network, they are processed step-wise by the NIDS. Each step that follows integrates the target network delays and responses. For target network response integration, TANTRA is executed on a proxy that controls the timing of the outgoing traffic, packet by packet. Therefore, packets can be adjusted in real time.
This represents a much more challenging setup, but it enables a real-world applicable execution, as TANTRA adjusts its approach according to frequent target network's response.

\subsection{Novel Defense Directions} \label{subsubsec:theoDefense}

For mitigation, we look to the principles of CV-based DeepFake detection.
DeepFake primarily relates to content, such as videos, generated by deep learning networks \cite{mirsky2020creation}. DeepFake algorithms can be used for malicious activities that may cause psychological, political, monetary, and physical harm. Given such implications, defenses have been developed using machine learning models to differentiate between real and fake videos (DeepFakes)~\cite{guera2018deepfake, gandhi2020adversarial}. 

A common approach used is to adjust the training procedure of existing anomaly detectors so that they are aware of the existence of DeepFakes. Similarly, we propose training three different ML classifiers (NIDS), namely logistic regression, Gaussian naive Bayes, and random forest, to learn to detect reshaped traffic. We hypothesize that increasing NIDS' awareness of reshaped characteristics  will enable them to distinguish between benign traffic and reshaped malicious traffic. More information on NIDS weakness is described in the Section \ref{sec:theorydefense}.

\section{Weaknesses of NIDS}\label{sec:theorydefense}

We observe that existing NIDSs aim to detect malicious traffic by searching for characteristics which are considered abnormal for benign traffic. This approach is used by all three state-of-the-art NIDSs examined in this study. Therefore, each NIDS is trained on benign traffic and searches for anomalies in network traffic to identify attacks.

For higher granularity in their prediction, all three of the systems perform their detection based on the feature space. Common practices involve using the AfterImage~\cite{mirsky2018kitsune} feature extractor, which achieves SOTA results, represents the industry standard~\cite{han2020practical, chalapathy2019deep, schneider2018high}, and is used in this work. Using AfterImage, nine basic problem-space attributes from each packet (MAC, IP, and protocol attributes for each source and destination, together with  the IP type, packet size, and timestamp) are used to extract features. 
Modifying one attribute affects all other attributes in the feature space. Therefore, changing the timestamp attribute affects the other attributes in the feature space. The extracted features then serve as input to the NIDS.
Therefore, TANTRA aims to exploit the fact that NIDSs are trained only on benign traffic and reshapes malicious traffic accordingly. We assess this weakness for each NIDS individually below. 

{\noindent{\textbf{Autoencoder (AE).}} AEs are neural networks that aim to reconstruct their input; in doing so, AEs attempt to characterize specific behavior of and patterns in the traffic, enabling them able to distinguish differences in high-dimensional space when abnormal input is introduced. }
The AE represents a function $f$ which aims to deconstruct and reconstruct an input. For any given input $x$ from the training set, $f(x$ aims at minimizing the reconstruction error RMSE. In the context of intrusion detection, the AE aims for the benign network packet, $b$, and its output $f(b)$, $f(b) - b = RMSE$ to be small. Similarly, it aims for a malicious network packet, $m$, and its output $f(m)$, $f(b) - b = RMSE$, to be large. In our case, we use the commonly used root mean square error (RMSE) as reconstruction error. 
\begin{equation}
RMSE(\vec{X}, \vec{Y}) = \sqrt{\frac{\sum_{i=1}^{n}\left ( x_{i} - y_{i} \right )^2}{n}}
\end{equation}

Our novel reshaping technique exploits this type of NIDS by modifying the malicious network traffic so it follows benign behavior.
Since it reshapes the timing characteristics of the attacker's malicious traffic, the attack will be perceived similarly to benign traffic by the AE. Thus, we hypothesize that the reconstruction error (RMSE) between $m$ and $f(m)$ will be both small and below a defined detection threshold.

\noindent\paragraph{\textbf{KitNET.}} Proposed by Mirsky et al.~\cite{mirsky2018kitsune}, KitNET is an unsupervised neural network composed of an ensemble of AEs. Using an ensemble of AEs provides the ability to use simpler AE architectures and lowers the computational complexity.
Therefore, like an AE, KitNET tries to characterize benign network traffic, determining $f(b) - b = RMSE$ that knows how to reconstruct benign network traffic $b$. 
We expect KitNET's reconstruction error of reshaped malicious traffic to be small, like that of AEs, and therefore likely to evade detection.

\noindent\paragraph{\textbf{Isolation Forest.}} The isolation forest (IF)~\cite{liu2008isolation} algorithm is an anomaly detection algorithm commonly used in related studies~\cite{ding2013anomaly, puggini2018enhanced, liu2012isolation}. IF creates clusters of  benign data; when encountering malicious input, the IF algorithm should isolate such inputs, based on their features, outside the clusters.
IF defines its anomaly score for a given example $x$ as:

\begin{equation}
s(x,n) = 2^{-\frac{E(h(x))}{c(n)}}
\end{equation}

where $E(h(x))$ is the average path length of observation, $x$, to a cluster (referred to as a tree); $c(n)$ is the average path length of unsuccessful searches in the trees; and, $n$, is the number of external nodes. An anomaly score close to one indicates an anomaly. Scores much smaller than 0.5 indicate normal observations, as the path's length to the current example, $x$, is smaller or equal to the average path length.
The theory behind the IF model includes the principle that anomalous examples behave differently than benign ones; thus, it will be easier to isolate anomalous examples than to detect examples which are typical for benign cases After reshaping the malicious traffic so it behaves similarly to benign traffic, we hypothesize that $E(h(x))$ will be close to the path length of benign examples, $c(n)$. Therefore, the reshaped malicious traffic should be similar to benign traffic and thus will have a similar average path length and evade detection.

\section{Evaluation}\label{sec:evaluation}
In this section, we evaluate TANTRA's effectiveness and mitigation. We first examine its baseline performance against state-of-the-art NIDSs following the setup used in prior research (Section \ref{sec:theorydefense}). Then, we optimize the attack by identifying the optimal LSTM window size and assess the attack's effectiveness in an end-to-end scenario with an active network connection. Finally, we evaluate the proposed defense technique as a mitigation against our novel evasion attack. 

\subsection{Experimental Setup}

\subsubsection{Datasets} \label{Datasets}

\begin{table*}[t]
\begin{center}
\caption{The datasets and utilization for experiments (time in minutes).}
\label{tab:datasets}
\resizebox{2\columnwidth}{!}{%
\begin{tabular}{cccccc}
\hline\hline 
Dataset & Attack Type & Total Recording & Benign NIDS Training & Benign LSTM Training & Malicious Testing \tabularnewline
\hline\hline
\multirow{6}{*}{Kitsune Dataset \cite{mirsky2018kitsune}} & Active Wiretap & 21.93 & 8.27 & 8.12 & 11.46\tabularnewline
 & MITM & 20.17 & 8.05 & 7.16 & 6.97\tabularnewline
 & Fuzzing & 28.95 & 12.96 & 10.91 & 14.58\tabularnewline
 & Mirai & 118.95 & 33.68 & 31.61 & 75.82\tabularnewline
 & SSDP Flood & 40.74 & 14.40 & 14.37 & 4.04\tabularnewline
 & SSL Renegotiation & 38.73 & 14.69 & 12.38 & 22.05\tabularnewline
\multirow{2}{*}{CIC IDS 2017 \cite{sharafaldin2018toward}} & Brute-Force & 61.1 & 14.43 & 0.23 & 4.15\tabularnewline
 & SQL Injection & 57.25 & 14.43 & 11.28 & 2.8\tabularnewline
\hline\hline
\end{tabular}
}
\end{center}
\end{table*}

We use the Kitsune dataset~\cite{mirsky2018kitsune} and intrusion detection evaluation dataset CIC-IDS2017 \cite{sharafaldin2018toward}. Both benchmarks have been used in related studies \cite{han2020practical, Rallying}. The datasets contain network traffic recordings of eight real network attacks, along with the benign and malicious labels for the network traffic packets.
The network traffic recordings are provided as PCAP files, which are retrieved in individual network setups. The overall length of the recordings ranges from 28.15 to 118.9 minutes, with an average of 59.35 minutes, and an average of 2,540,013 packets. Table \ref{tab:datasets} presents (1) the total recording time, (2) the portion used to train the NIDS (benign traffic only), (3) the portion used to train TANTRA's LSTM (benign traffic only), and (4) the portion used to evaluate TANTRA (malicious traffic only). 

\subsubsection{NIDS}
TANTRA is evaluated on three NIDSs that cover both traditional machine learning and deep learning approaches. 
After the traffic is reshaped by TANTRA in the problem space, it is extracted to the feature space where it serves as input to the NIDS. As done in other studies, we use the AfterImage feature extractor which converts the PCAP files into features that can serve as NIDS input.

    \noindent\paragraph{\textbf{Autoencoder.}} The AE NIDS consists of two parts - an encoder and a decoder. We use a convolutional neural network with a two-layer encoder ($100x64$ neurons), latent space of 32 neurons, and two-layer decoder architecture ($64x100$ neurons). For training, the Adam optimizer is utilized with a learning rate of 0.001, and the mean squared error (MSE) is used as the loss criterion. The Keras framework is employed \cite{chollet2015keras}.
    \noindent\paragraph{\textbf{KitNET.}} We follow KitNET NIDS' open-source code with the default maximum autoencoder size of 10 in the ensemble layer, a learning rate of 0.1, and a hidden ratio of 0.75. 
    We set the mapping and training phase parameters according to the work presenting Kitsune. RMSE is used as the loss criterion.  
    \noindent\paragraph{\textbf{Isolation Forest.}} The Isolation Forest NIDS has 100 base estimators in the ensemble, 
    with the threshold set to auto contamination, and the random state equal to zero and default anomaly score function~\cite{scikit-learn}.

\subsubsection{Evaluation Metrics}\label{Metrics}

The detection rate (DR) of the NIDS for detecting malicious traffic is used to evaluate TANTRA's performance. The changes in the DR as a result of our evasion attack are used to quantify TANTRA's effectiveness; this is also referred to as recall.
The DR metric has typically been used to showcase either a large DR when proposing a new NIDS~\cite{folorunso2016nids, li2020improving, mirsky2018kitsune} or a low DR when proposing a new attack technique~\cite{adv03, IDSGAN, DLNIDS}.

\subsection{TANTRA Baseline Evaluation} \label{Transferability}

In this section, we provide a baseline evaluation, following prior work in which all of the traffic is prepared on the attacker's side and evaluated against the NIDS directly. This evaluation setup disregards any potential target network behavior, such as network delays and responses, which may influence successful evasion. Although past studies have been unable to demonstrate an effective evasion attack in an end-to-end scenario, we perform the first step of the evaluation using their setup for comparison and method optimization purposes.

We train an LSTM with the window size set at three, which also utilizes an Adam optimizer with $0.001$ learning rate, and the mean squared error (MSE) loss function. 
For each attack, an LSTM is trained on benign traffic that matches the traffic on the targeted on which the attack is performed. 
On average, the benign traffic (training set) for each attack consists of 296,501 packets from the network recordings, with a minimum of 73,875 packets.
We evaluate the DR of the three NIDSs before and after reshaping the attack network traffic.

\begin{table*}[t]
\caption{NIDS detection rate before and after TANTRA (in \%).}
\label{tab:models_comparison}
\centering
\resizebox{1.68\columnwidth}{!}{ 
\begin{tabular}{ccccccc|cc}

\hline\hline
\multirow{2}{*}{Attack Type} & \multicolumn{2}{c}{AE} & \multicolumn{2}{c}{KitNET} & \multicolumn{2}{c|}{IF} & \multicolumn{2}{c}{KitNET (After)}\tabularnewline
 & Before & After$\downarrow$ & Before & After$\downarrow$ & Before & After$\downarrow$ & Related Work \cite{han2020practical}$\downarrow$ & TANTRA$\downarrow$ \tabularnewline
\hline\hline
Active Wiretap & $98.03$ & $0.00$ & $99.02$ & $0.00$ & $68.73$ & $7.29$ & N/A & N/A\tabularnewline
MitM & $23.79$ & $0.00$ & $78.20$ & $1.37$ & $1.47$ & $0.00$ & N/A & N/A\tabularnewline
Fuzzing & $67.53$ & $0.97$ & $92.37$ & $0.00$ & $49.28$ & $0.83$ & $1.31$ & $0.00$\tabularnewline
Mirai & $88.94$ & $0.00$ & $100.00$ & $0.00$ & $0.83$ & $0.42$ & $0.58$ & $0.00$\tabularnewline
SSDP Flood & $71.94$ & $10.22$ & $99.97$ & $0.00$ & $99.94$ & $0.01$ & $21.47$ & $0.00$\tabularnewline
SSL Renegotiation & $89.34$ & $1.19$ & $39.19$ & $0.00$ & $6.80$ & $0.00$ & N/A & N/A\tabularnewline
Brute-Force & $25.50$ & $0.00$ & $22.79$ & $0.00$ & $87.94$ & $9.34$ & $28.18$ & $0.00$\tabularnewline
SQL Injection & $23.81$ & $0.00$ & $33.33$ & $0.00$ & $100.00$ & $17.91$ & N/A & N/A\tabularnewline
\hline 
\textbf{Average} & \textbf{$61.11$}&\textbf{$1.55$}& \textbf{$70.61$} & \textbf{$0.17$} & \textbf{$51.87$} & \textbf{$4.48$}& \textbf{$12.89$}& \textbf{$0.00$}\tabularnewline
\hline\hline 
\end{tabular}}\end{table*}

The results presented in Table \ref{tab:models_comparison} show that after applying our evasion attack to each malicious dataset, the network traffic in each dataset is able to successfully evade the NIDS (lowering the DR). After reshaping the attack traffic using our evasion attack, the DR decreases on average from 61.19\% to 2.06\%.

Without using TANTRA, the AE NIDS is able to detect most network attacks with an average DR of 61.11\%. However, after reshaping the attack, TANTRA almost completely conceals the attack, reducing the average DR on the AE to 1.55\%. While the DR of most attacks is reduced to below 1\%, the SSDP flood attack is an exception. In this case, the initial detection rate of 71.94\% is reduced to 10.22\%, which is ten times higher than the average DR after reshaping for the other attacks. One possible explanation for this behavior is that the SSDP attack is an attack that contains several phases, with responses between some of the phases. Since all of the attack network traffic is generated before being sent to the target network, this evaluation setting prohibits integrating the network response time in each phase. Thereby, this illustrates the drawback of the evaluation setting used in past work and the relevance of evaluating an end-to-end scenario.

\begin{figure}[t]
\begin{center}
\includegraphics[width=0.98\columnwidth]{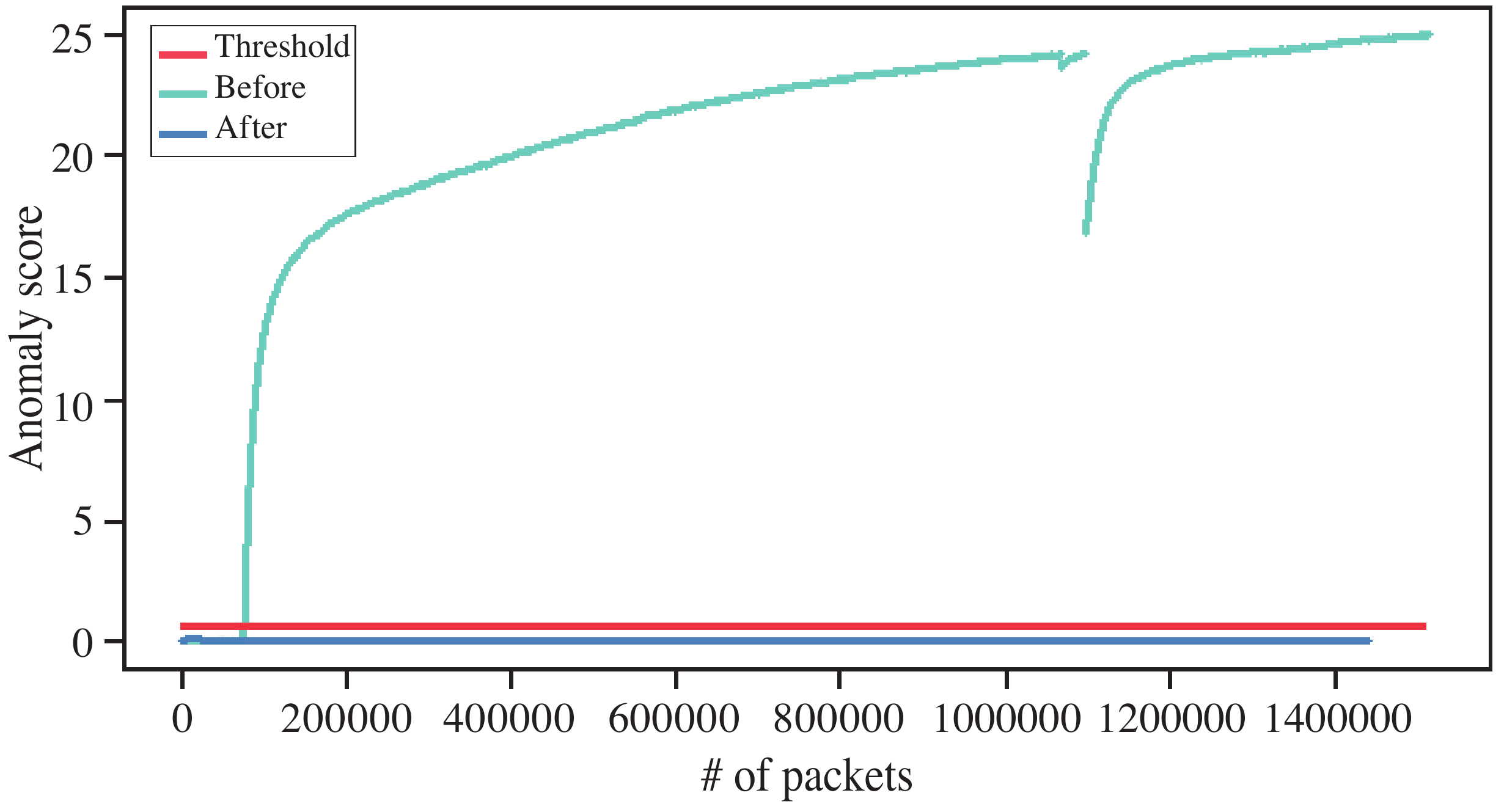}
\caption{AE NIDS anomaly score for Mirai attack.}
\label{fig:AE_Mirai}
\end{center}
\end{figure}

Figure \ref{fig:AE_Mirai} presents an example of the behavior of the AE when classifying malicious traffic. In the example, we present the anomaly score for the Mirai attack, before and after using TANTRA.
The x-axis represents the number of processed network packets in a sequence, and the y-axis represents the anomaly score. The anomaly score before TANTRA (turquoise) is far above the detection threshold (red line). After TANTRA, the anomaly score for the Mirai attack is below the threshold for all network traffic. Therefore it can be seen that the AE is unable to detect the malicious behavior, which results in a DR of 0\%.

The greatest impact of the evasion attack is observed when the KitNET NIDS is used. Before reshaping, KitNET performs best, with an average DR of 70.61\%. After reshaping, the average DR decreases to 0.17\%, the lowest DR among the three NIDSs.
Another interesting finding is that TANTRA achieves a perfect score for seven of the eight network attacks examined.
One possible explanation for the significant results when using KitNET is that the autoencoder ensemble is likely to be the best at characterizing benign behavior. As TANTRA aims to exploit benign behavior, it reshapes the malicious traffic so it \textbf{behaves} like benign traffic, thereby evading detection, aligned with our theoretical assessment (see Section~\ref{sec:theorydefense}). 

Prior research has also used KitNET NIDS for evaluation~\cite{han2020practical}, which serves as baseline comparison. Here, the performance is evaluated on four attacks, as shown in Table \ref{tab:models_comparison}. The results of the study show that the attacks presented in prior research are able to decrease detection by the KitNET NIDS by an average of 12.89\% with DRs ranging from 0.58\% to 28.18\%. For the evaluated attacks, TANTRA decreases detection to 0.00\% for all attacks. Other prior research used white-box settings which are difficult to compare directly to TANTRA. Therefore we present additional qualitative comparison in Section \ref{sec:comparison}.

\begin{figure}[t]
\begin{center}
\includegraphics[width=\columnwidth]{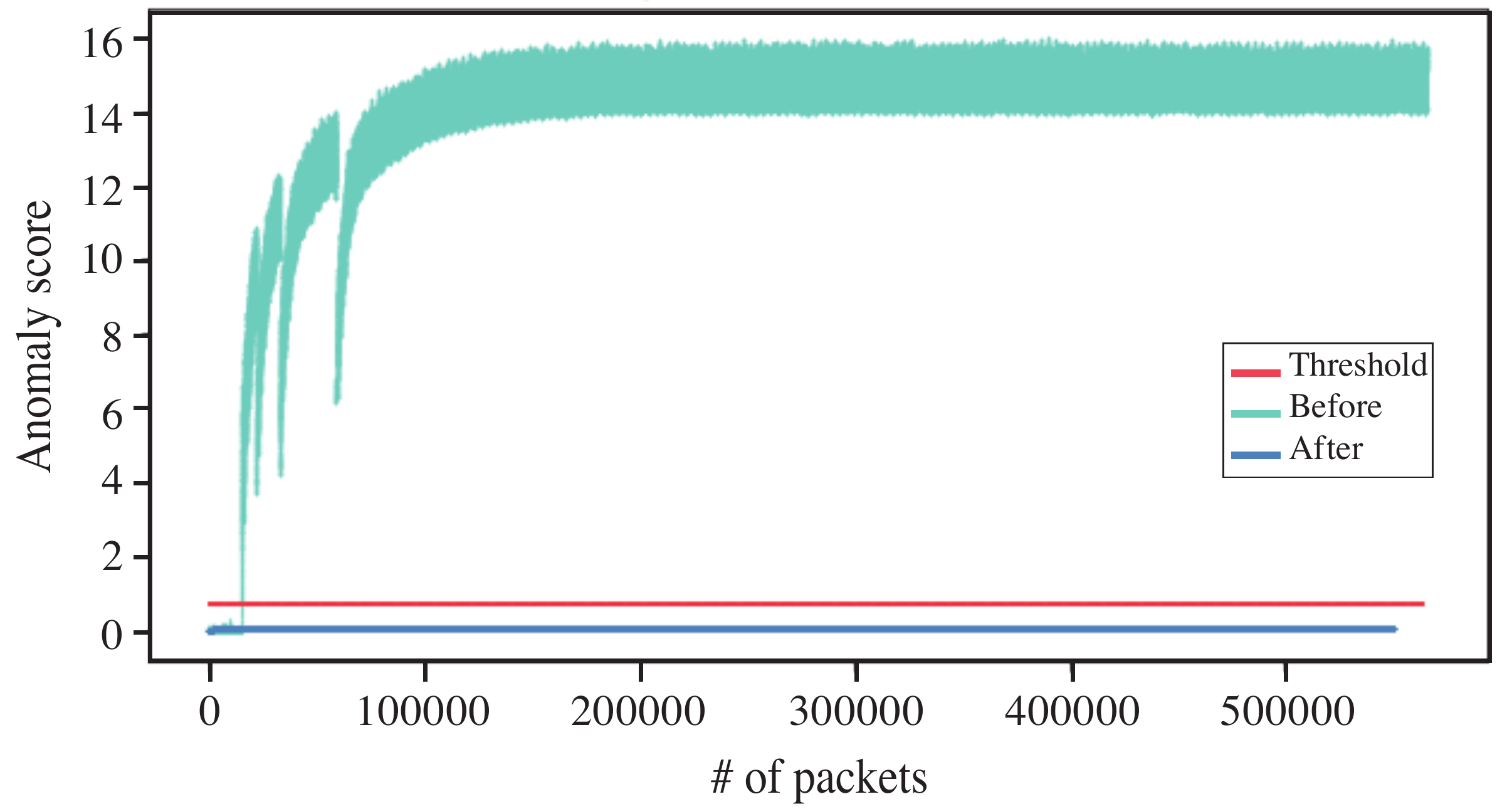}
\caption{KitNET NIDS anomaly scores for SSDP flood attack.}
\label{fig:Kit_SSDP}
\end{center}
\end{figure}

Figure \ref{fig:Kit_SSDP} provides a visual example for KitNET. This figure presents the anomaly score for the SSDP flood attack. Again, the turquoise lines, which represent the malicious traffic before reshaping, are far above the detection threshold. However, after reshaping, the NIDS fails to detect the traffic as malicious, as represented by the dark blue lines below the red threshold. 

The lowest DR before TANTRA is applied is obtained by the IF NIDS. Before reshaping, the IF has an average DR of 51.87\%, obtaining a DR below 10\% with three of the eight attacks. However, after reshaping, the IF has the highest DR (an average DR of 4.48\%) of the three intrusion detection systems. This DR stems from the relatively higher detection rate this NIDS obtained for three attacks after TANTRA was used, namely SQL injection with a DR of 17.91\%, brute-force with 9.34\%, and active wiretap with 7.29\%. The DR for the other five attacks is below 1.0\%. 

\begin{figure}[t]
\begin{center}
\includegraphics[width=\columnwidth]{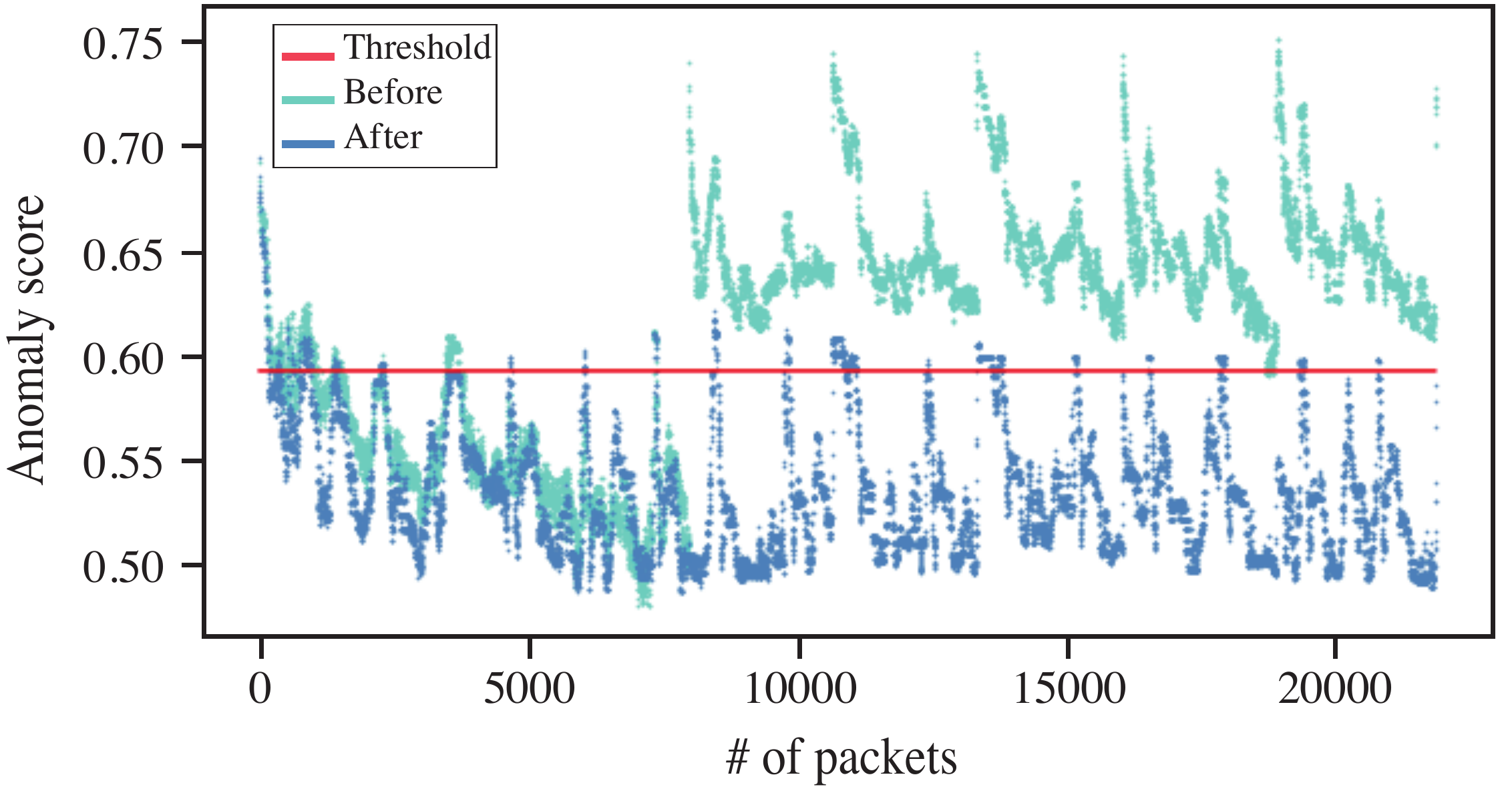}
\caption{IF NIDS anomaly score for MITM attack.}
\label{fig:IF_ARP}
\end{center}
\end{figure}

Figure \ref{fig:IF_ARP} presents the anomaly scores of the IF for the MITM attack, before and after TANTRA. The results support our claim that many malicious packets are different from benign examples before reshaping. After reshaping, most malicious packets are labeled benign by the IF model.

\subsubsection{LSTM Optimization}\label{subsec:WSOpt}
While improving the attacks' ability to evade detection, for most attacks the use of TANTRA did not result in a DR of 0\%.  For example using TANTRA for the active wiretap attack against the IF NIDS results in a DR of 7.29\%. When using TANTRA for the SSDP flood attack against the AE, the DR is 10.22\%. Therefore, in this section, we explore how TANTRA can be optimized, to ensure that it is capable of fully evading any type of state-of-the-art NIDS, regardless of the type of attack being reshaped.

Given the problem space's limited attributes, we turned to the LSTM hyperparameters for optimization. Here, we focus on the window size, which controls how much information about past predictions influences the current one. For example, with a window size of three, only the last three prediction outcomes are integrated into the prediction, while with a window size of 10, the last 10 predictions are utilized, and so on. 

There is no perfect solution for the optimal window size, as this varies depending on the attack and input size. Olah et al. \cite{olah2015understanding} demonstrated this issue in the natural language processing (NLP) field with a translation task. Given the sentence, "I grew up in France. I speak fluent \underline{French}," our goal is to predict the last word. If we only take the previous three words into account, "I speak fluent \_\_\_\_," our model will most likely not predict the right word, whereas if we use the previous four words or more, the probability of the model to be correct is much higher, as the word "France" is now part of the history. 

Factors to consider regarding the appropriate window size are the overall length of the input (analogous to the length of a sentence) and the context dependencies between the input elements (analogous to the context dependencies between words in a sentence). For TANTRA's LSTM, the input elements are network packet properties. The length of the network traffic varies depending on the attack. In the datasets, an active wiretap attack consists of roughly 500,000 packets, while the brute-force attack has only 1,507 packets. Therefore, a different window size may be needed depending on the attack.

The results presented in Table \ref{tab:window_optimization} show, that increasing the window size improves the attack's evasion capabilities, decreasing the average DR to 2.98\% for WS=3, 0.88\% for WS=50, and 0.10\% for WS=150. 
We find that enlarging the WS reduces the DR further so that it is close to or equal to 0\% for all network attacks. For example, attacks with fewer packets, such as the brute-force attack with 1,507 packets, can fully evade the IF NIDS when the window size is enlarged from three to 50. 

For other attacks against the IF NIDS, e.g., the active wiretap attack, which has 500,000 network packets, the increase in window size has to be even larger. Here, increasing the window size to 150 results in almost perfect evasion, with a DR of 0.79\%. Similar behavior is observed for the other NIDSs. For example, reshaping the SSDP flood attack (1,500,000 packets) with a window size of 150 against the AE NIDS results in a 100\% success rate (0\% DR).
Overall, we find that all attacks are undetected when the window size is set at 150. Given the fact that an attacker usually does not know which NIDS is used by the target network, it thus makes sense to use an LSTM window size of 150.

\begin{table}[t]
\begin{center}
\caption{NIDS DR with increasing LSTM window size (in \%).}
\label{tab:window_optimization}
\resizebox{1.0\columnwidth}{!}{%
\begin{tabular}{ccccc}
\hline\hline 
\multirow{2}{*}{NIDS/Attack Type} & \multirow{2}{*}{Before} & \multicolumn{3}{c}{After$\downarrow$}\tabularnewline
 &  & WS=3 & WS=50 & WS=150\tabularnewline
\hline\hline 
AE/Active Wiretap & 98.03 & 0.00 & 0.79 & 0.00\tabularnewline
AE/SSDP Flood & 71.94 & 10.22 & 0.00 & 0.00\tabularnewline
AE/Brute-Force & 25.50 & 0.00 & 0.00 & 0.00\tabularnewline
KitNET/Active Wiretap & 90.02 & 0.00 & 0.00 & 0.20\tabularnewline
KitNET/SSDP Flood & 99.97 & 0.00 & 0.00 & 0.00\tabularnewline
KitNET/Brute-Force & 22.79 & 0.00 & 0.00 & 0.00\tabularnewline
IF/Active Wiretap & 69.73 & 7.29 & 7.09 & 0.73\tabularnewline
IF/SSDP Flood & 99.94 & 0.01 & 0.01 & 0.00\tabularnewline
IF/Brute-Force & 87.94 & 9.34 & 0.00 & 0.00\tabularnewline
\hline 
\textbf{Average} & \textbf{73.98} & \textbf{2.98} & \textbf{0.88} & \textbf{0.10}\tabularnewline
\hline\hline 
\end{tabular}
}

\end{center}
\end{table}

\begin{tcolorbox}[size=title,colback=white]
{\textbf{Impact Summary}: Having optimized TANTRA, all eight attacks remain undetected against all three NIDSs, decreasing the average DR to 0.01\%. Compared to prior studies, this demonstrates an up to 28.18\% improvement and a 12.89\% average improvement (Table \ref{tab:models_comparison}). Furthermore, the attacks functionality is preserved as only attack network traffic timestamps are reshaped.}
\end{tcolorbox}

\subsection{End-to-End Evaluation}\label{Real}
In the previous section, we examined the effectiveness of TANTRA when reshaping all of the attack network traffic and evaluated the attack against an NIDS; this enables us to perform a comparison to related work. One of TANTRA's key strengths, however, is its end-to-end capability, which allows the attacks to be done using an active target network connection, as would be the case for a real-world scenario. TANTRA faces a more difficult challenge in this setup, as it has to reshape attack network traffic step by step, according to the target network's delays and responses in real time.

To perform this evaluation, we (1) set up a target network with a KitNET NIDS, (2) train the timestamp generating LSTM on benign network traffic, (3) reshape the MITM malicious attack traffic using the trained LSTM, and (4) evaluate TANTRA against the NIDS. 
We then set up a proxy using the Scapy \cite{biondi2010scapy} library to send each packet with the specific timestamp generated by the LSTM model. In this way, network traffic is sent to the target network while adjusting the timestamps in real time. 

\begin{figure}[t]
  \includegraphics[width=\columnwidth]{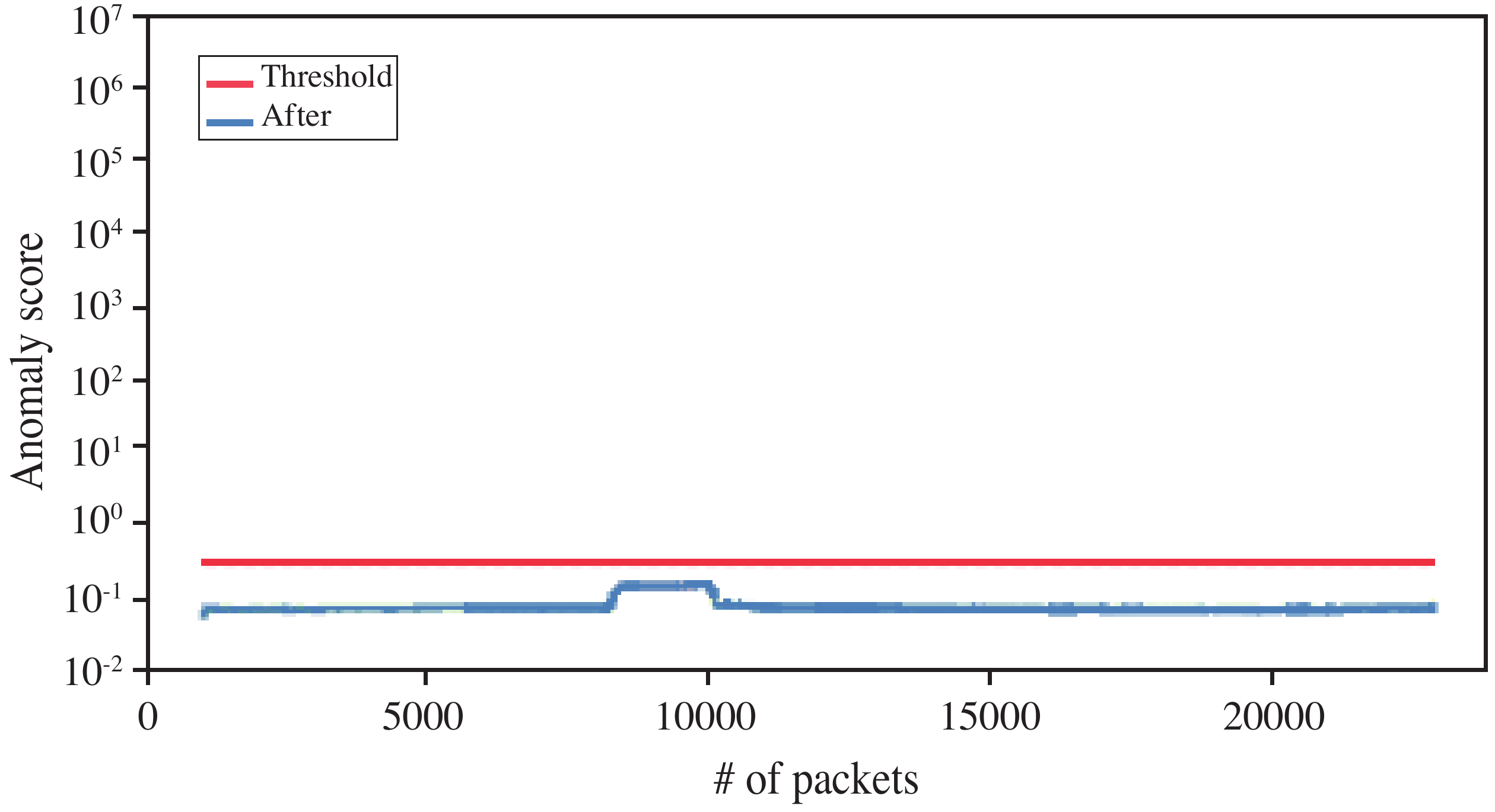}
  \caption{KitNET NIDS anomaly scores for MITM attack in end-to-end scenario.}
    \label{fig:Real_time}
        
\end{figure}

Figure \ref{fig:Real_time} shows KitNET's anomaly scores when processing TANTRA's reshaped MITM network traffic. The results indicate that the reshaping results in full evasion throughout the active connection.

One interesting observation is that after the first step, from the ~348,000 packet onward, the anomaly score increases slightly. This could be due to additional network delay and response activity. This may illustrate the relevance of an independent end-to-end evaluation for validating the performance under real-world conditions. Despite the slight increases in the anomaly score, the attack's scores remain below the NIDS detection threshold. Therefore, we consider TANTRA the first viable approach capable of overcoming the challenges of an end-to-end scenario, while remaining fully undetected and retaining the functionality of the attack network packet contents. 

In future work, this scenario can be evaluated for other types of intrusion attacks. Furthermore, other target networks can be explored, enabling different types of proxy to be tested for intrusion. In addition, other types of NIDSs can be integrated. In addition, by optimizing the window size, TANTRA can be fine-tuned for specific attack, network, and NIDS characteristics, to obtain the greatest impact. Each of these settings can be evaluated using the general approach presented in this paper.

\begin{tcolorbox}[size=title,colback=white]
{\textbf{Impact Summary}: 
While additional difficulty is imposed by target network delays and responses, by using TANTRA, attacks can evade detection throughout the entire intrusion process. The observed anomaly scores on KitNET show the impact of network responses, which resulted in a spike in the score. However, TANTRA's ability to integrate such responses in the reshaping process resulted in a 0\% DR for the evaluated attack.
}
\end{tcolorbox}

\subsection{Mitigating TANTRA}\label{subsec:defenseEval}

To identify leading directions for mitigating TANTRA, we propose a new way of training an NIDS. We use the methodology presented in Section \ref{subsubsec:theoDefense}, but instead of training on benign behavior for the detection of unknown anomalies, we train on both benign traffic and reshaped malicious traffic for the purpose of distinguishing between them.
The suggested mitigation is evaluated on three supervised-learning ML-based NIDSs with the following settings:

\begin{itemize}
    \item \textbf{Logistic Regression (LR).} We use an logistic regression NIDS following scikit-learn \cite{scikit-learn} default hyperparameter setting, where we do not execute normalization (normalize=False), and we do not force the coefficients to be positive (positive=False). Furthermore, maximum iterations of convergence are set to 100.
    \item \textbf{Gaussian Naive Bayes (NB).} The Gaussian Naive Bayes NIDS also follows the default settings by the scikit-learn, where the portion of the largest variance of all features that is added to variances for calculation stability is 1e-9 (var\_smoothing=1e-09). 
    \item \textbf{Random Forest (RF).} For the Random Forest NIDS, we use 100 estimators and the Gini criterion to measure the impurity of the estimator splits. We do not define a maximum depth and set the minimum sample splits to 2. Furthermore, bootstrap samples are used to build the trees.

\end{itemize}

The NIDSs are evaluated as described in Section \ref{Transferability}. We use both small and large packet size attacks, namely active wiretap, MITM, fuzzing, and SSDP flood attacks. We evaluate the NIDSs using cross-validation. Benign traffic, along with three reshaped attacks, is used for training. Benign and reshaped malicious traffic is trained in equal proportion. The remaining fourth attack is used for testing.

The results presented in Table \ref{tab:defense_results} show that the defense technique achieves a total average DR of 45.93\% at the cost of a 20.97\% false positive rate (FPR). Both, the NB NIDS and RF NIDS are able to detect almost all reshaped malicious traffic for the MITM and SSDP flood attacks. However, the NB NIDS is the only one that achieves such performance with a 0\% FPR. In contrast, the RF NIDS has an FPR of 17.73\%. 

The least effective NIDS seems to be the LR NIDS whose average FPR is 6.53 percentage points higher than the average DR. A possible explanation for such behavior may be that the NB and RF NIDSs focus on dealing with continuous data, which may help find connections between different features. On the other hand, the LR NIDS aims to build a linear function to differentiate between classes, which may be more difficult in our case.

Interestingly, none of the NIDSs were able to detect the active wiretap or fuzzing attacks after TANTRA was used. One reason for that may be that both attacks are designed for execution on a router, while MITM and SSDP flood attacks are designed for execution inside a connection in the target network, as shown by Mirsky et al. \cite{mirsky2018kitsune}. Therefore, the attack design differentiates the four attacks, which may cause the NIDS training process to be less effective on two of them.

Overall, the defense technique seems to be highly effective for some attacks, while other attacks remain undetected. While not yet fully effective, the proposed mitigation provides a promising direction for addressing TANTRA.

\noindent\paragraph{\textbf{Challenges and Discussion.}}
To further mitigate TANTRA, general and versatile defense techniques must be identified. So far, changing the training paradigm has shown promising results for two out of the four evaluated attacks. One of the main challenges is to train NIDSs so that they are able to detect any kind of reshaped malicious traffic. Our proposed defense technique uses reshaped malicious traffic from TANRA's methodology to train the NIDS specifically for such traffic. However, as evasion attacks evolve, the trained NIDS may become outdated and unable to cope with the unfamiliar new reshaping methods. Hence, there is a need for a general defense technique that can detect reshaped malicious traffic, even when the method for reshaping is unknown to the defender. 

The second challenge that mitigation faces has to do with the NIDS used to perform detection. In this mitigation evaluation, different NIDSs than those used when evaluating our attack. The reason is the autoencoder's compatibility with non-binary decisions. Therefore, two different classes with four decision outcomes would require an individual AE, which then could be combined in an ensemble, confirming or disregarding one another; we propose pursuing this direction in future research. 

Finally, the identified mitigation directions could be integrated into existing NIDSs which are trained solely on benign traffic. In such a case, the detection systems could work in a combined fashion in which the first one provides an anomaly score, and the second one is used to predict whether or not the input is a malicious or benign sample. However, the optimal scenario would require training the NIDSs, which are aware of reshaped malicious traffic, against all variants of evasion and maintaining such detection systems with the latest attacks, similar to traditional malware classifiers such as antivirus software.

\begin{table}[]
\begin{center}
\caption{NIDS DR following novel defense technique.}
\label{tab:defense_results}
\vspace{10pt}
% \resizebox{0.9\columnwidth}{!}{

\begin{tabular}{cccc}
\hline \hline 
NIDS & Attack Type & DR$\downarrow$ & FPR$\downarrow$ \tabularnewline
\hline \hline 
\multirow{5}{*}{Logistic Regression} & Active Wiretap & 0.00 & 50.70\tabularnewline
 & MITM & 100.00 & 33.60\tabularnewline
 & Fuzzing & 0.00 & 83.00\tabularnewline
 & SSDP Flood & 54.60 & 13.40\tabularnewline
\cline{2-4} \cline{3-4} \cline{4-4} 
 & Average & 38.65 & 45.18\tabularnewline
\hline 
\multirow{5}{*}{Gaussian Naive Bayes} & Active Wiretap & 0.00 & 0.00\tabularnewline
 & MITM & 100.00 & 0.00\tabularnewline
 & Fuzzing & 0.00 & 0.00\tabularnewline
 & SSDP Flood & 100.00 & 0.00\tabularnewline
\cline{2-4} \cline{3-4} \cline{4-4} 
 & Average & 50.00 & 0.00\tabularnewline
\hline 
\multirow{5}{*}{Random Forest} & Active Wiretap & 0.00 & 0.00\tabularnewline
 & MITM & 100.00 & 39.90\tabularnewline
 & Fuzzing & 0.00 & 0.00\tabularnewline
 & SSDP Flood & 96.50 & 31.00\tabularnewline
\cline{2-4} \cline{3-4} \cline{4-4} 

 & Average & 49.13 & 17.73\tabularnewline
\hline \hline 

\multicolumn{2}{c}{\textbf{Total Average}} & \textbf{45.93} & \textbf{20.97}\tabularnewline

\hline \hline 

\end{tabular}

\end{center}

\end{table}

\section{Related Work}\label{sec:comparison}
In this section, we present related work and compare the methods proposed in those studies to TANTRA, in terms of the preparation required before the attack is executed, the attack's execution, the ability to remain undetected, the impact of the attack and other ways of mitigation.

The majority of related studies have focused on reshaping the feature space. Lin et al. \cite{IDSGAN} generated adversarial attacks using a unique type of generative adversarial network (GAN), namely IDSGAN. Yang et al. used internal information about the detection system to generate adversarial examples. Wang et al. \cite{DLNIDS} used a multi-layer perceptron (MLP), which was trained using the original training dataset, to generate adversarial examples using both the FGSM and JSMA adversarial attack methods. Warzyński and Kołaczek \cite{IDSVol} attempted to compromise an NIDS based on a surrogate model. While Clement et al. \cite{Rallying} assumed that the adversary has direct knowledge of the target DL NIDS, allowing them to generate input for the deep learning model directly.
The work closest to ours is the study performed by Han et al. \cite{han2020practical}. The authors did not use the feature space; instead, they proposed an evasion attack to bypass an NIDS, by combining the use of NIDS behavior analysis and dummy packets to reshape the traffic.

\noindent\paragraph{\textbf{Preparation.}} Most related work assumed to have access to the targeted system's feature extraction process~\cite{Rallying, adv04, IDSGAN, IDSVol, DLNIDS, adv03}. With regard to the preparation required for the attack, the approaches presented in other papers mainly relied on training a separate DL model that is used to perform an adversarial attack. In other cases, there was a need to obtain access to the targeted classifier's input and evaluate its output~\cite{Rallying, adv04, adv03}. While capable of bypassing  an NIDS, the required knowledge makes these attacks impractical. 
    
Gaining access to NIDS behavior relies on threat models which are both difficult to build and require information and knowledge that is difficult to obtain, which questions the attacks' value when compared to the benefits of evading the NIDS.
Hat et al.trained a GAN to produce reshaped traffic by integrating dummy packets into the attack network traffic. To do so, all of the traffic must be generated~\cite{han2020practical} . This approach neglects the issue of integrating target network delays or responses, which makes it inapplicable in an end-to-end scenario under the proposed setup.
TANTRA's preparation consists of training an LSTM on benign traffic obtained from the target network.
In contrast to related studies, knowledge about the NIDS is not needed, meaning that TANTRA is a black-box evasion attack.

\noindent  \paragraph{\textbf{Execution \& Performance.}} When comparing execution and performance, it is crucial to distinguish between white-box and black-box evasion attacks. White-box evasion attacks assume barely obtainable knowledge which may increase their performance. This performance, however, is rarely achievable in a real-world setting. Black-box evasion attacks are better suited for direct comparison to TANTRA.

In the case of white-box attacks, Clement et al. achieved 100\% success for integrity attacks, while Lin et al. achieved similar results, as they decreased the DR to at most 1.56\%. Rigaki et al. were able to decrease the targeted NIDS' accuracy by 0.25\%, on average, while Wang et al. were able to decrease the DR from an average of 56.128\% to 20.0\% for most combinations of adversarial and network attacks.

For black-box attacks, Han et al. used similar to our setup, the Mirai, fuzzing, SSDP flood, and brute-force attacks and evaluated the attacks after reshaping on a KitNET NIDS. Thereby the authors' approach decreased the average detection rate to 12.89\%. In contrast, the use of TANTRA decreased the DR to 0\%. This makes it possible to fully evade those attacks against a KitNET NIDS; we also validate our method on two other state-of-the-art NIDSs (AE and IF NIDSs). Furthermore, TANTRA is able to obtain similar performance for the other attacks evaluated in this work.

This comparison shows that TANTRA outperforms the related black-box evasion attacks and even outperforms white-box evasion attacks which have  far greater knowledge on the NIDS.

Execution-wise, in comparison to related work, TANTRA simply reshapes attack network traffic using an LSTM and the timestamp attribute. This may limit the flexibility in reshaping the traffic, however it guarantees functional attack network traffic.
Moreover, TANTRA is able to interact with the target network in each step of the attack in real-time, allowing it to be effective in more difficult settings.

\noindent \paragraph{\textbf{Impact.}} The impact of bypassing an NIDS can be great. For example, it can allow an attacker to take control of a targeted network, listen to network traffic, or deny service. The attacks presented in other studies that use the feature space rarely have an impact on the target network. First, such attacks require knowledge on the NIDS that is very difficult to obtain. Second, even if the knowledge is obtained, the attack is likely to lose its functionality when the reshaped feature space is transformed back to the problem space (Section \ref{sec:mainchallenges}). These studies may serve as the basis for the development of more impactful attacks if a means for such reconstruction is developed in the future. 

On the other hand, the black-box evasion attacks presented in Han et al.~\cite{han2020practical} may have impact. Regarding the attack presented in that paper, there is a risk that the attack will be inexecutable given the use of dummy packages or be detected because of the inability to integrate network responses or delays. But there is still a chance that the approach will be successful for some other cases in a real-world setting. 
In contrast, TANTRA's effectiveness was demonstrated in a full end-to-end scenario with an active network connection, where it was 100\% successful in evading detection by the KitNET NIDS for various attacks. 

\noindent\paragraph{\textbf{Mitigation.}} Some mitigation has been proposed by prior research. One of the proposed defense technique is adversarial feature reduction~\cite{han2020practical}. Here, dimensions of features are reduced to those which provide the NIDS the greatest insight into distinguishing between malicious and benign packets. While this may be effective against the attack presented in their paper, it is not the case for TANTRA. Since TANTRA's approach focuses on emulating benign behavior, adversarial feature reduction will have difficulties in defending against and difficulties to detect the reshaped malicious traffic of TANTRA. 

Other studies trained networks on adversarial robustness, exposing the NIDS to adversarial attacks during training, in order to guide anomaly detection~\cite{advrob}. Similar behavior is observed in the emerging area of out-of-distribution detection. In this case, the system is exposed to outliers, called outlier exposure~\cite{oe_defense}, which is then integrated into the training procedure using a special loss function. As a result, the system behaves differently to the outliers which are detected by a drop in prediction accuracy or overall prediction entropy. 

Overall, with the help of the defense technique presented in this paper, we identified that exposing the system to reshaped malicious traffic helps improve detection for some attacks. However, the proposed defense technique still requires optimization for being considered a generic approach to evading evasion attacks.

\section{Conclusion}\label{sec:conclusion}
In this work, we presented TANTRA, the first end-to-end evasion attack that can evade all state-of-the-art NIDSs with a 99.99\% success rate. 
We showed that by simply reshaping attack network traffic using the timestamp attribute, TANTRA outperforms related work's evasion attacks by up to 28.18\%. Furthermore, the attack was capable of successfully evading the NIDS when executed in a black-box setting. When evaluated under an active network connection, TANTRA bypassed KitNET, one of the most advanced NIDSs, and is the first evasion attack that does not affect the attack packets' content. Therefore, the attack's functionality is retained, allowing the attack to impact the target network. 

We also presented a promising mitigation direction that involves training a NIDS with both benign and reshaped traffic. While not yet fully effective against all types of attacks, this defense technique shows to be a promising direction and is worth pursuing in future research, as a means of defending against the TANTRA threat presented in this paper.  

{\footnotesize \bibliographystyle{acm}
\bibliography{main}}

\end{document}